\documentclass[11pt]{article}

\usepackage{amsmath,amsfonts,amsbsy,amssymb,array,accents,dsfont}
\usepackage{enumerate,array,latexsym,graphicx,mathrsfs,verbatim,psfrag}
\usepackage{bm} 
\usepackage[normalem]{ulem}
\usepackage{multirow}

\usepackage{booktabs} 
\usepackage[usenames]{color}
\usepackage[utf8]{inputenc}
\usepackage[english]{babel}
\usepackage{fancybox}

\usepackage{datetime}

\usepackage[nosort]{cite}
\usepackage{chngpage} 

\usepackage{enumitem}
\setlist{itemsep=0pt}

\usepackage[colorlinks=true,      linkcolor=darkblue,      urlcolor=darkblue,      
            filecolor=darkblue,      citecolor=darkblue,       pdfstartview=FitH,     
						pdfpagemode=UseNone,      bookmarksopen=true]{hyperref}  
\usepackage[all]{hypcap}     

\newcommand{\captionfonts}{\small}
\makeatletter  
\long\def\@makecaption#1#2{%
  \vskip\abovecaptionskip
  \sbox\@tempboxa{{\captionfonts #1: #2}}%
 \ifdim \wd\@tempboxa >\hsize
    {\captionfonts #1: #2\par}
  \else
    \hbox to\hsize{\hfil\box\@tempboxa\hfil}%
  \fi
  \vskip\belowcaptionskip}
\makeatother   

\DeclareMathSymbol{\medhatsym}{\mathord}{largesymbols}{"62} 

\DeclareMathSymbol{\medtildesym}{\mathord}{largesymbols}{"65}

\newcommand{\comm}[1]{} 

\def\IC{\mathbb{C}}

\def\IR{\mathbb{R}}

\def\({\left(}
\def\){\right)}
\def\[{\left[}
\def\]{\right]}

\def\coeff#1#2{{\textstyle \frac{#1}{#2}}}

\def\One{{\hbox{ 1\kern-.8mm l}}}

\def\barray{\begin{array}}
\def\earray{\end{array}}
\def\be{\begin{equation}}
\def\ee{\end{equation}}
\def\bea{\begin{eqnarray}}
\def\eea{\end{eqnarray}}
\def\bal{\begin{align}}
\def\eal{\end{align}}

\numberwithin{equation}{section} 

\makeatletter
\g@addto@macro\bfseries{\boldmath}
\makeatother

\definecolor{cardinal}{rgb}{0.6,0,0}
\definecolor{darkgreen}{rgb}{0,0.4,0}
\definecolor{purple}{rgb}{0.5, 0, 0.5}
\definecolor{golden}{rgb}{0.92, 0.7, 0}
\definecolor{midnight}{rgb}{0, 0, 0.5}
\definecolor{darkblue}{rgb}{0, 0, 0.8}

\def\IC{\mathbb{C}}
\def\Neql#1{{\cal N}\!=\!{#1}}
\def\coeff#1#2{\relax{\textstyle {#1 \over #2}}\displaystyle}

\def\IR{\mathds{R}}
\def\ZZ{\mathds{Z}}

\def\cB{{\cal B}}

\def\cD{{\cal D}}

\def\cK{{\cal K}}
\def\cL{{\cal L}}

\def\cR{{\cal R}}

\begin{document}


\begin{flushright}
%
%
\end{flushright}

\vspace{14mm}

\begin{center}

{\huge \bf{The Structure of BPS Equations }} \medskip \\
{\huge \bf{for Ambi-polar Microstate Geometries}}\medskip


\vspace{13mm}

\centerline{{\bf Alexander Tyukov$^1$, Robert Walker$^{1}$ and Nicholas P. Warner$^{1,2}$}}
\bigskip
\bigskip
\vspace{1mm}

\centerline{$^1$\,Department of Physics and Astronomy,}
\centerline{University of Southern California,} \centerline{Los
Angeles, CA 90089-0484, USA}
\bigskip
\centerline{$^2$\,Department of Mathematics,}
\centerline{University of Southern California,} \centerline{Los
Angeles, CA 90089, USA}

\vspace{4mm}

{\small\upshape\ttfamily  ~tyukov @ usc.edu,  walkerra @ usc.edu, ~warner @ usc.edu} \\

\vspace{10mm}
 
\textsc{Abstract}

\begin{adjustwidth}{17mm}{17mm} 
 %
\vspace{3mm}
\noindent
Ambi-polar metrics, defined so as to allow the signature to change from $+4$ to $-4$  across hypersurfaces, are a mainstay in the construction of BPS microstate geometries.  This paper elucidates the cohomology of these spaces so as to simplify greatly the construction of infinite families of fluctuating ``harmonic'' magnetic fluxes. It is argued that such fluxes should come from scalar, harmonic pre-potentials whose source loci are holomorphic divisors. This insight is obtained by exploring the K\"ahler structure of ambi-polar Gibbons-Hawking spaces and it is shown that differentiating the pre-potentials with respect to K\"ahler  moduli yields solutions to the BPS equations for the electric potentials sourced by the magnetic fluxes. This suggests that harmonic  analysis on ambi-polar spaces has a novel, and an extremely rich structure, that is deeply intertwined with the BPS equations.  We illustrate our results using a family of two-centered solutions.

%
\end{adjustwidth}

\end{center}


\thispagestyle{empty}

\newpage


\baselineskip=14pt
\parskip=2pt

\tableofcontents


\baselineskip=15pt
\parskip=3pt

\section{Introduction}
\label{Sect:Intro}
 
The microstate geometry program seeks to find smooth, horizonless geometries that look like black holes until one is very close to the horizon scale.  Instead of having a horizon and singularity, microstate geometries are stably causal solitons that cap off smoothly at some finite red-shift.  In (3+1)-dimensions there are ``no-go'' theorems showing that such solitonic solutions do not exist in Einstein-Maxwell theory.  However, with more spatial dimensions, there are now well-established techniques that use non-trivial cohomological fluxes to support solitonic geometries against collapse and the formation of horizons.  Indeed, the old ``no-go'' theorems have been updated \cite{Gibbons:2013tqa} to show that such non-trivial cohomological fluxes represent essentially the {\it only} mechanism that can support  smooth, solitonic geometries that look like black holes.

In practice the construction of such microstate geometries is much simpler when they are supersymmetric (or BPS) and this is largely because a major part of the underlying BPS equations are linear  \cite{Bena:2004de,Bena:2011dd}. As a result, there are now systematic procedures for constructing large families of BPS microstate geometries  \cite{Bena:2005va,Berglund:2005vb,Bena:2007kg,Bena:2015bea}. There are also some limited systematic procedures for non-BPS microstate geometries but these are far more complicated in their implementation  \cite{Bossard:2014ola,Bena:2015drs,Bena:2016dbw,Bossard:2017vii}. 

The standard BPS black hole is sourced by singular electric charges with the black-hole mass locked onto the values of the charges.  One can  obtain microstate geometries by making a geometric transition in which these electric charges are replaced by smooth, cohomological magnetic fluxes that generate the charges via Chern-Simons interactions.   These magnetic fluxes have to be dual to non-trivial homology cycles, and so the geometric transition necessarily involves  blowing-up, or resolving, the singular sources into cycles threaded by flux.  The geometry is then supported against collapse by these fluxes and a horizon never forms. 

We now have a huge  variety of BPS microstate geometries and, in particular, there are families of {\it scaling} microstate geometries that have  arbitrarily-deep AdS throats, before they cap off in smooth geometries just above where the horizon of the corresponding black-hole would be.    Because of the  AdS throats, holographic field theory has been used to great effect to study the precise microstate structure that is encoded in such geometries.  (See \cite{Bena:2015bea,Bena:2016agb,Bena:2016ypk,Bena:2017geu, Bena:2017xbt} for recent developments.)  In addition, the essential role of  geometric transitions can be understood in terms of emergence of new infra-red phases in the conformal field theory underpinning the black-hole microstate structure (see, for example, \cite{Bena:2013dka}).

The physics of microstate geometries is now very well-developed, but the underlying mathematical structure of the BPS equations still remains somewhat mysterious because of the essential role of ambi-polar metrics\footnote{Ambi-polar metrics are allowed change signature from $+4$ to $-4$ across hypersurfaces, but are otherwise smooth geometries.  This idea will be discussed and defined more carefully in later sections of this paper.} in realizing geometric transitions.  In this paper we will examine this more closely and elucidate some of the properties and structure of the BPS equations.  In particular, we will show that the  cohomological fluxes can be obtained from scalar pre-potentials and that part of the solution to the BPS equations can be obtained from the K\"ahler deformations of these pre-potentials.  For simplicity, we will restrict ourselves to microstate geometries in five dimensions.  There are probably very interesting, parallel results for six-dimensional microstate geometries.

Our broader purpose is two-fold.  For physics, we wish to streamline the construction of BPS solutions so that more complicated, fluctuating, multi-centered geometries might fall within reach of computations.  Obtaining such solutions is not simply an academic exercise, but will help us understand the holographic interpretation of such geometries and elucidate their role in describing black-hole microstate structure.  For mathematics, we hope to frame the cohomology of ambi-polar geometries in a more precise manner and show how the standard cohomological lore of Riemannian, K\"ahler geometry becomes much richer in ambi-polar geometries.

In Section \ref{Sect:BPS-5d} we briefly review the BPS equations and their structure in five dimensions. In particular, we discuss ambi-polar metrics and clarify what we mean by a ``harmonic form'' on an ambi-polar space.    In Section \ref{Sect:ComplexGH} we consider  Gibbons-Hawking (GH) metrics and  their formulation as K\"ahler geometries. For Riemannian GH metrics, the first parts of Section  \ref{Sect:ComplexGH}  represents a fairly standard review, however we actually allow the metrics to be ambi-polar. This is used to motivate the conjecture that the harmonic forms  have pre-potentials.  We then discuss the relationship between moduli, Lichnerowicz modes and harmonic forms and show how the appropriately covariant derivatives of pre-potentials provide solutions to the second layer of the BPS equations.  In Section \ref{Sect:TwoCenter} we show how our conjectured structure of ambi-polar cohomology  works beautifully for two-centered solutions.  In  Section \ref{Sect:Conc}  we summarize our results and make some further speculations about the structure of cohomology on ambi-polar spaces and how the structure of BPS solutions may be further simplified and developed.

\section{The BPS structure in five dimensions } 
\label{Sect:BPS-5d}

In five-dimensions, the starting point for constructing microstate geometries is a generic stationary space-time over a four-dimensional spatial base, $\cB$:
\begin{equation}
ds_5^2 ~\equiv~ - Z^{-2}  (dt+k)^2 ~+~  Z \, ds_\cB^2 \,, \qquad  ds_\cB^2 ~\equiv~ h_{\mu \nu}dx^\mu dx^\nu \,.
\label{fivemetric}
\end{equation}
This is used as a background for five-dimensional $\Neql 2$ supergravity coupled to $n$ vector multiplets.  For the solution to preserve one supersymmetry, the  four-dimensional metric, $ds_\cB^2$, must be hyper-K\"ahler.  Including the gravi-photon, the theory contains  $(n+1)$ electromagnetic fields, whose (local) vector potentials are decomposed according to:
\begin{equation}
A^{(I)}  ~=~ -  Z_I^{-1} \, (dt +k) ~+~ B^{(I)} \,,
\label{Apots}
\end{equation}
where the $Z_I^{-1}$ are the electrostatic potentials and the $B^{(I)}$ are (local) magnetic components purely on $\cB$.   The magnetic field strengths are then defined by
\begin{equation}
\Theta^{(I)} ~\equiv~ d B^{(I)}   \,.
\label{Thetadefn}
\end{equation}
The electromagnetic sector also involves a completely symmetric, cubic coupling, $C_{IJK}$, that satisfies a number of  constraints that we will not detail here (see, for example, \cite{Gutowski:2004yv,Gauntlett:2004qy}).

In the geometry (\ref{fivemetric}), supersymmetry imposes a constraint on the metric function, $Z$:
\begin{equation}
Z ~\equiv~ \big(\coeff{1}{6}\, C^{IJK} Z_I Z_J Z_K\big)^{1/3}   \,,
\label{Zconstraint}
\end{equation}
which relates the mass to the charges.  Supersymmetry of the whole solution is then guaranteed if one then solves the linear BPS system on $\cB$:
\begin{align}
 \Theta^{(I)}  &~=~  \star_4 \, \Theta^{(I)} \label{BPSeqn:1} \,, \\
 \nabla^2  Z_I &~=~  \coeff{1}{2} \, \star_4 \, C_{IJK}  (\Theta^{(J)} \wedge  \Theta^{(K)})  \label{BPSeqn:2} \,, \\
 dk ~+~  \star_4 dk &~=~  Z_I \,  \Theta^{(I)}\,.
\label{BPSeqn:3}
\end{align}
The simplest, generic\footnote{Here ``generic'' connotes the fact  that one can use dualities to rotate a single object with a generic charge vector into the ``STU'' model.} example of this is the ``STU'' model in which there are two vector multiples and $C_{IJK}  =|\epsilon_{IJK}|$. We will need a slight generalization of this in Section \ref{Sect:TwoCenter}.

Since the $\Theta^{(I)}$ arise from  local potentials, (\ref{Thetadefn}), the $\Theta^{(I)}$ are closed and so the first equation, (\ref{BPSeqn:1}) implies that the $\Theta^{(I)}$ are also co-closed.  If these magnetic fields are also smooth, then, according to the standard definition, they are harmonic on $\cB$. In Riemannian geometries the non-trivial, harmonic  fluxes must be cohomological duals of the $2$-cycles on $\cB$.

The base geometry is  four-dimensional and hyper-K\"ahler.  If one imposes that this geometry is smooth, Riemannian and asymptotic to $\IR^4$ then the only such geometry is flat $\IR^4$.  It therefore originally seemed that there could not be any interesting microstate geometries that look like black holes in $(4+1)$-dimensions. However, it was evident from \cite{Giusto:2004kj} that the four-dimensional base need not be Riemannian.  Indeed the base can be ``ambi-polar,'' in that its signature can flip from $+4$ to $-4$ as one passes across ``evanescent ergosurfaces'' in $\cB$.  The crucial point is that the five-dimensional, Lorentzian metric can still be smooth despite that apparent pathology of the metric on $\cB$.   To achieve this, the function $Z$ must change sign as one passes across an evanescent ergosurface.  Indeed, $Z$ passes through a pole on these surfaces and so the Killing vector, $\frac{\partial}{\partial t}$, is time-like on either side of the surface and null on the surface. This is why such  surfaces  are called evanescent ergosurfaces.  One can show that, despite the pole in $Z$, the complete metric is smooth across evanescent ergosurfaces \cite{Bena:2007kg}.

This observation opened up a raft of new possibilities that have been extensively exploited over the last 13 years and have led to the rich families of known microstate geometries.  Despite this, the systematic mathematical structure of ambi-polar geometries is only now beginning to be explored    \cite{Hitchin1, Biquard1,Niehoff:2016gbi}.  For physicists, the best working definition of ``regularity'' on ambi-polar metrics is if the resulting five-dimensional, Lorentzian metric, (\ref{fivemetric}), is  smooth and stably causal.  Equivalently, the four-dimensional spatial geometry obtained from constant time-slices of (\ref{fivemetric}) (setting  $dt =0$) must be smooth and Riemannian.  
It is useful to denote the smooth, Riemannian metric on $\cB$ by 
\begin{equation}
\widetilde{ds}_4^2 ~\equiv~ - Z^{-2} \, k^2 ~+~  Z \, ds_\cB^2  \,.
\label{reg4met}
\end{equation}
Note that this metric on $\cB$ is not hyper-K\"ahler. Indeed there has been very little analysis of the geometric properties of this class of metrics. 

Similarly, regularity of Maxwell fields is defined in the five-dimensional metric, or on its constant time slices.  That is,  Maxwell fields obtained by setting $dt=0$ in (\ref{Apots}): 
\begin{equation}
\tilde A^{(I)}  ~=~ -  Z_I^{-1} \, k  ~+~ B^{(I)} \,, \qquad \widetilde F^{(I)}  ~\equiv~ d \tilde A^{(I)}  
\label{regMax}  \,.
\end{equation}
are required to be smooth on $\cB$ . The Maxwell fields $\widetilde F^{(I)}$ are certainly closed but are neither  self-dual nor co-closed.  Again, there has been very little analysis of the  properties of this class of Maxwell fields. 

Herein lies the abiding mathematical mystery of  the BPS equations:  They are simple, linear and have an ``upper-triangular'' form only when expressed in terms of an intrinsically singular geometry: the ambi-polar base metric.  This metric is hyper-K\"ahler and the magnetic fluxes are harmonic only in the ambi-polar base.  Yet, regularity or smoothness is only determined in a constant-time slicing of the five-dimensional metric and Maxwell fields.  In particular, the metric and harmonic forms on the ambi-polar space are singular on the evanescent ergosurfaces.

Up until the results presented in \cite{Bena:2017geu}, it was thought that despite these differences, the  harmonic $2$-forms on ambi-polar spaces could be constructed by transferring standard results from Riemannian hyper-K\"ahler manifolds.  Indeed, the $2$-cycles on $\cB$ are regular in the geometry defined by  (\ref{reg4met}) and, while the canonical harmonic $2$-forms, $\Theta$,  are singular on $\cB$, they can be ``improved'' to obtain smooth Maxwell fields given by (\ref{regMax}).  Moreover, the periods of these improved magnetic Maxwell fields are readily computed and can be seen to provide a dual basis to the homology.  In this sense, the standard ideas of smooth cohomology and Hodge and Poicar\'e duality carry across to ambi-polar spaces.  

What changes between Riemannian and ambi-polar spaces, and became apparent in  \cite{Bena:2017geu},  is that the canonical harmonic representatives, $\Theta$, of the cohomology are far from the complete set of harmonic 2-forms: indeed, there are infinite families of harmonic $2$-forms.  This has its origins in the absence of a positive definite $\cL^{(2)}$-norm on forms in the ambi-polar metric, $ds_\cB^2$.  However, in practice, it means that there are infinitely many closed and co-closed, self-dual forms, $\Theta$,  whose improvements defined via (\ref{regMax}), are smooth and of finite norm on $\cB$ in the metric $\widetilde{ds}_4^2$.   Henceforth we will define $\Theta$'s to be {\it harmonic} on an ambi-polar geometry if they are closed, co-closed and the complete system  (\ref{BPSeqn:1})--(\ref{BPSeqn:3})  produces a smooth, improved Maxwell field  (\ref{regMax}).

From the point of view of Riemannian cohomology using $\widetilde{ds}_4^2$, the infinite families of $\Theta$'s boil down to allowing a vast number of cohomologically exact pieces to be  added to $\widetilde F^{(I)}$ in (\ref{regMax}). Such infinite families are cohomologically trivial in that they do not modify the periods of the canonical magnetic fluxes.  

For the physics of BPS microstate geometries, these infinite families of ``harmonic'' $\Theta$'s are hugely important.  They represent fluctuating electromagnetic fields  on the $2$-cycles and they can, through $Z$ and $k$, produce fluctuations in the shapes of the cycles.  Indeed, in  \cite{Bena:2017geu} it was shown that these modes represent momentum waves and angular-momentum excitations of the MSW string.  

On a more practical level, what is being said is that there are infinite families of cohomologically exact modes that can produce rich families of physical solutions to the BPS equations (\ref{BPSeqn:1})--(\ref{BPSeqn:3}) and that one misses all these families if one makes the simple mistake of assuming that because $\Theta$ is  closed and co-closed and self-dual, it must be restricted to the finite family of ``canonical'' cohomological fluxes that are drawn from the corresponding Riemannian geometries. 

One of our purposes in this paper will be to study, in more detail, the properties of such harmonic $2$-forms in ambi-polar geometries. 
 
\section{The complex structure of GH geometries} 
\label{Sect:ComplexGH}

\subsection{The metric}
\label{ss:metric}

We consider the standard form of Gibbons-Hawking geometries:
\begin{equation}
ds_4^2 ~=~ V^{-1} \, \big( d\psi + \vec{A} \cdot d\vec{y}\big)^2  ~+~
 V\, (d\vec{y} \cdot d\vec{y}) \,,
\label{GHmetric}
\end{equation}
with 
\begin{equation}
\vec \nabla \times \vec A ~=~ \vec \nabla V\,.
\label{AVreln}
\end{equation}
We also use the standard set of frames
\begin{equation}
\hat e^1~=~ V^{-{1\over 2}}\, (d\psi + A) \,,  \qquad \hat e^{a+1} ~=~ V^{1\over 2}\, dy^a \,, \quad a=1,2,3 \,.
\label{GHframes}
\end{equation}
With (\ref{AVreln}), and this choice of frame, the Riemann tensor is self-dual.  The metric is thus hyper-K\"ahler. 

The standard form of the Gibbons-Hawking metrics involves taking
\begin{equation}
 V ~=~  \sum_{j=1}^N \,  {q_j  \over r_j} \,,
\label{Vform}
\end{equation}
where  $r_j \equiv |\vec{y}-\vec{y}^{(j)}|$ for some $\vec{y}^{(j)} \in \IR^3$.  Regularity requires one to take $q_j \in \ZZ$ and, for the metric to be Riemannian, one must require $q_j > 0$. However, our focus here are the ambi-polar metrics and so we will allow the $q_j$ to be negative.  

For simplicity, we will also take the $\vec{y}^{(j)}$ to be co-linear,  and choose axes so that these points all lie along the $z \equiv y_3$-axis at points $z=a_j$.  We introduce polar coordinates, $(\rho, \phi)$ perpendicular to the $z$-axis: 
\begin{equation}
y_1 + i \, y_2 ~=~  \rho \, e^{i \phi} \,, \qquad r_j ~=~ \sqrt{\rho^2 + (z-a_j)^2} \,.
\label{polars}
\end{equation}
The metric has two $U(1)$ isometries defined by the Killing vectors $\frac{\partial}{\partial \psi}$ and $\frac{\partial}{\partial \phi}\,$.

It is also convenient to introduce sets of spherical polar coordinates, $(r_j, \theta_j, \phi)$, with origins at $\vec{y}^{(j)}$:
\begin{equation}
\rho  ~=~   r_j \,  \sin \theta_j \,,  \qquad (z-a_j)  ~=~   r_j \,  \cos \theta_j \,.
\label{sphpolars}
\end{equation}
In particular, the vector potential, $A$, is given by  
\begin{equation}
A ~=~ A_\phi \, d \phi \,, \qquad  A_\phi ~\equiv~ \sum_{j=1}^N \, q_j \, \cos \theta_j  \,.
\label{Aform}
\end{equation}
%

\subsection{The K\"ahler structure}
\label{ss:Kahler}

Define the following natural bases for the self-dual and anti-self-dual  two-forms:
\begin{equation}
\Omega_\pm^{(a)} ~\equiv~ \hat e^1  \wedge \hat
e^{a+1} ~\pm~ \coeff{1}{2}\, \epsilon_{abc}\,\hat e^{b+1}  \wedge
\hat e^{c+1} \,, \qquad a =1,2,3\,.\
\label{twoforms}
\end{equation}
The two-forms, $\Omega_-^{(a)}$, are anti-self-dual and harmonic and  define the  hyper-K\"ahler  structure on the base.   We are going to select one of these as a K\"ahler form and use it to define a complex structure.  Indeed, we take:
\begin{equation}
J ~=~ \Omega_{-}^{(3)} ~=~ (d\psi +A)  \wedge dz ~-~ V \, dy_1 \wedge dy_2~=~ (d\psi +A)  \wedge dz ~-~V \,  \rho \, d\rho \wedge d\phi \,.
\label{Kahlerform}
\end{equation}

Since the metric has two $U(1)$ isometries, it is easy to construct the complex coordinates explicitly using the moment maps.  The  holomorphic $1$-forms obtained from the Killing vectors are 
\begin{equation}
\omega_\psi ~=~V^{-1}\, (d\psi + A) ~+~i\, dz   \,, \qquad \omega_\phi ~=~    V^{-1}\,  A_\phi  \, (d\psi + A)  ~+~ V \rho^2 d \phi   ~+~ i\, \Big[A_\phi \, dz  ~+~  V \rho \, d \rho  \Big] \,.
\label{holforms1}
\end{equation}
As a result, we have the following exact holomorphic forms:
\begin{align}
\rho^{-1}\, d \rho    ~-~   i \,  d \phi   ~=~  &   -i\,(\rho^{2} \, V)^{-1}\, (\omega_\phi - A_\phi \, \omega_\psi)   \,, \\
 -i \,d \psi     ~+~ \sum_{j=1}^N  q_j \, \bigg(\frac{d z}{r_j}  - \cos\theta_j \, \frac{d \rho}{\rho}  \bigg)   ~=~  &   -i\,V \, \omega_\psi ~+~i\,A_\phi \, (\rho^2 \, V)^{-1}\, (\omega_\phi - A_\phi \, \omega_\psi)  
\label{holforms2}
\end{align}
which can be integrated to give the following complex coordinates\footnote{The sign discrepancy between (\ref{polars}) and (\ref{z12defn}) is due to to the canonical choice of complex structure (\ref{Kahlerform}). To remedy this one would need to use use the anti-canonical choice $J\to-\Omega_{-}^{(3)}$.}:
\begin{equation}
\zeta_1 ~=~  e^{-i \psi}  \,  \prod_{j=1}^N \bigg( \frac{\cos \coeff{1}{2} \theta_j }{ \sin \coeff{1}{2} \theta_j} \bigg)^{q_{j}}  \,, \qquad  \zeta_2 ~=~  \rho\, e^{-i \phi}     \,.
\label{z12defn}
\end{equation}
One can then re-write the metric in hermitian form:
\begin{equation}
ds_4^2 ~=~ V^{-1} \, \bigg| \frac{d \zeta_1}{ \zeta_1} ~+~A_\phi\, \frac{d \zeta_2}{ \zeta_2}  \bigg|^2  ~+~
 V\,  \big| d\zeta_2  \big|^2  \,.
\label{hermGHmetric}
\end{equation}
The easiest way to find the K\"ahler potential, $\cK$, is to work with the real coordinates and use the identities
\begin{equation}
J ~=~  d C \,, \qquad C_\mu ~=~ {J_\mu}^\rho \, \partial_ \rho \, \cK\,.
\label{Jpot}
\end{equation}
for some vector potential, $C$. One easily finds that 
\begin{equation}
C ~=~ - z \, (d \psi + A_\phi \, d \phi) ~-~ (F - z \, A_\phi)\, d \phi  \,, \qquad  F ~\equiv~  \sum_{j=1}^N \,  q_j  \, r_j   \,.
\label{Cdefn}
\end{equation}
Contracting with ${J_\mu}^\rho$ and integrating leads to
\begin{equation}
\cK ~=~ -  \sum_{j=1}^N \,    q_j\, \bigg[\, r_j ~+~ a_j \log \bigg( \frac{\cos \coeff{1}{2} \theta_j }{\sin \coeff{1}{2} \theta_j} \bigg) \bigg]  \,.
\label{Kpot}
\end{equation}
One can interpret this simply as the sum of the K\"ahler potentials in $\IC^{2(N+1)}$ pushed through the hyper-K\"ahler quotient construction.

\subsection{Harmonic forms and pre-potentials}
\label{ss:harmonic}

The  $2$-forms, $\Theta_{\mu \nu}$, that we seek are self-dual.  This means that, considered as matrices, ${\Theta_\mu}^ \nu$ and ${J_\mu}^ \nu$ must commute because $J$ is anti-self-dual. It follows that 
\begin{equation}
{J_\mu}^\rho \, {J_\nu}^\sigma \, \Theta_{\rho \sigma}~=~   \Theta_{\mu \nu}  \,,
\label{JJTheta}
\end{equation}
and therefore self-dual $2$-forms must be $(1,1)$-forms in the complex structure defined by $J$.  The fact that the $\Theta$'s are also harmonic means that they can be used to make K\"ahler deformations that preserve the Ricci-flatness of the metric.  Such deformed metrics therefore still have $SU(2)$ holonomy and are therefore necessarily hyper-K\"ahler.   In Riemannian GH metrics, these deformations correspond to moving the GH points, $y^{(j)}$, in directions parallel to the $z$-axis. We will return to this in Section  \ref{ss:compensators}.  For now, we simply note that, at linear order, this means that one can define the deformed K\"ahler form via 
\begin{equation}
\tilde J ~=~  J ~+~ \epsilon\, \Theta  \,,
\label{Jdeform}
\end{equation}
for some infinitessimal parameter, $\epsilon$.  Since $\tilde J$ must have a K\"ahler potential of the form $\tilde\cK = \cK + \epsilon \Phi$, for some function, $\Phi$, it follows that $\Theta$ must have a pre-potential defined by 
\begin{equation}
\Theta   ~=~    d\, \big( \big({J_\mu}^\rho \, \partial_ \rho \, \Phi \big) \,dx^\mu \big) \,.
\label{Prepot}
\end{equation}
Conversely, if one starts from the expression (\ref{Prepot}) as an Ansatz for $\Theta$, one finds that the resulting $(1,1)$ form is self-dual if and only if $\Phi$ is a harmonic function:
\begin{equation}
d\, \star_4 d \, \Phi  ~=~  0   \,.
\label{harmPhi}
\end{equation}
Thus, at least in Riemannian geometries, the harmonic $(1,1)$-forms can be obtained from harmonic pre-potentials on the GH manifold.

The canonical set of harmonic $2$-forms on a GH space are obtained by transferring the standard results for Riemannian geometries over to the more-general ambi-polar geometries.  Explicitly, one has:
\begin{equation}
\Theta^{(I)} ~ \equiv~ - \sum_{a=1}^3 \, \big(\partial_a \big( V^{-1}\, K^{(I)}  \big)\big) \,
\Omega_+^{(a)} \,,
\label{harmtwoform}
\end{equation}
where
\begin{equation}
 K^{(I)}  ~ =~   \sum_{i=1}^N \,  {k^{(I)}_i  \over r_i }  \,,
\label{Kharms}
\end{equation}
for some constant parameters $k^{(I)}_i$.  They are self-dual by construction and they are closed because the $K^{(I)}$ and $V$ are harmonic.  Regularity at $r_i \to 0$ is also guaranteed by the fact that the  $K^{(I)} $ are only singular when $V$ is singular. The $\Theta$'s are, however, singular on the evanescent ergosurfaces defined by $V=0$.

The vector potentials for these $\Theta^{(I)}$ are:
\begin{equation}
B^{(I)}  ~\equiv~ \frac{K^{(I)}}{V}  \, (d\psi ~+~ A) ~+~  \vec{\xi}^{(I)} \cdot  d \vec y \,,
\label{Bpot}
\end{equation}
where
\begin{equation}
\vec  \nabla \times \vec \xi^{(I)}  ~=~ - \vec \nabla K^{(I)} \,.
\label{xidefn}
\end{equation}
Once again, it is elementary to integrate ${J_\mu}^\rho  B^{(I)}_\rho d x^\mu$ to obtain the pre-potentials 
\begin{equation}
\Phi^{(I)}  ~=~  \sum_{i=1}^N \,    k^{(I)}_i\,  \log \bigg( \frac{\cos \coeff{1}{2} \theta_i }{\sin \coeff{1}{2} \theta_i} \bigg) \,.
\label{PhiIres}
\end{equation}

At first sight there seems to be an extremely simple relationship between (\ref{PhiIres}) and (\ref{Kpot}) in which one differentiates (\ref{Kpot}) with respect to the locations, $a_j$, of the GH points.  This  simplicity is an illusion because  the argument of the logarithm also  depends implicitly on the $a_j$.  We will return to this in Section  \ref{ss:NiceRes} and examine the derivatives of pre-potentials with respect to moduli.  

If one restricts one's attention to Riemannian GH metrics, then everything we have described has a rigorous mathematical underpinning.  The  $\Theta^{(I)}$ given in (\ref{harmtwoform}) yield an $(N-1)$-dimensional\footnote{Choosing $K^{(I)} = V$ yields a trivial $\Theta^{(I)}$.} basis of the harmonic $2$-forms  representing the cohomology.   The proof of many of the mathematical details rests heavily on the fact that there is a well-defined $\cL^{(2)}$-norm on forms and that the $\Theta^{(I)}$ have finite norm.  When the metric is ambi-polar, this is no longer true. However, as a practical matter it has long been known  \cite{Bena:2005va,Berglund:2005vb,Bena:2007kg} that many of these results carry directly across to ambi-polar metrics and their regular harmonic $2$-forms.    On ambi-polar spaces, the harmonic $2$-forms, like those in (\ref{harmtwoform}),  are necessarily singular on the evanescent ergosurfaces, defined by setting $V = 0$, but they are still harmonic under the definition given earlier for ambi-polar spaces.   The new element arising out of \cite{Bena:2017geu} is that this new definition also allows infinite families of fluctuating harmonic $2$-forms.

We are therefore going to take a ``leap of faith'' and assume that the broader, infinite families of ``harmonic'' $(1,1)$-forms that are important on ambi-polar backgrounds still have harmonic pre-potentials.   The obvious way to attempt a proof of this conjecture is to work within patches of the ambi-polar metric in which it is actually Riemannian and construct the pre-potentials within such patches. The issue then comes at the transitions between patches and especially on regions that include the evanescent ergosurfaces.  The stitching  can probably be achieved by going to the Riemannian metric,  (\ref{reg4met}), and the improved Maxwell fields (\ref{regMax}).  Rather than attempt this intricate mathematical analysis to justify our conjecture, we will take the more pragmatic view that it seems to work, at least for two-centered solutions.  We will describe this in detail in Section \ref{Sect:TwoCenter}, where we will also see that there is a very interesting  gauge choice  that  transfers the singular loci of the harmonic pre-potentials between the Riemannian patches.  

Another  very interesting open question  is whether these infinite families of infinitessimal variations of the K\"ahler potential can be integrated up to produce infinite families of hyper-K\"ahler moduli for ambi-polar hyper-K\"ahler metrics.  We will however take a first, putative step in this direction by examining the behavior of the pre-potentials as one varies the standard set of GH moduli.  This leads to a rather revealing insight into the role of the second BPS equation, (\ref{BPSeqn:2}).

\subsection{Compensators and covariant derivatives on the moduli space}
\label{ss:compensators}

If one makes a linear perturbation, $\delta g_{\mu \nu}$, of a Ricci flat metric and imposes the condition that the perturbed metric is also Ricci-flat to linear order, then the one finds that $\delta g_{\mu \nu}$ must be a zero-mode of a second-order linear operator, known as the Lichnerowicz operator.  On a K\"ahler manifold, smooth, Ricci-flat  K\"ahler perturbations can be obtained from the harmonic $(1,1)$ forms, $\Theta$, via 
\begin{equation}
\delta g_{\mu \nu} ~=~ \coeff{1}{2} \, \big( {J_\mu}^\rho \, \Theta_{\rho \nu} ~+~{J_\nu}^\rho  \, \Theta_{\rho \mu  } \big)   \,.
\label{LichModes}
\end{equation}
On GH spaces, there is no need to symmetrize because of the self-duality properties of $J$ and  $\Theta$.

Naively, if one  has a family of Ricci-flat metrics that depend upon a parameter, $a$, then one would expect  $\delta g_{\mu \nu} =  \partial_a g_{\mu \nu}$ to be a Lichnerowicz mode and related to harmonic forms, $\Theta$, via (\ref{LichModes}).  This is almost true.  One must, however, remember that metric components depend on coordinate systems and smoothness of a metric deformation may only be defined up to a compensating infinitessimal change of coordinates.   That is, given a Ricci-flat family of metrics depending on some parameters, $a_j$, the variations with respect to $a_j$ will only be smooth if the variations are combined with a suitably chosen infinitessimal coordinate change, $x^\mu \to x^\mu + Y_{(j)}^\mu$, for some vector field $Y_{(j)}^\mu$.  

Explicitly, the smooth Ricci flat metric variations are given by:
\begin{equation}
\delta_j g_{\mu \nu} ~=~\partial_{a_j}  g_{\mu \nu} ~+~ \cL_{Y_{(j)}} \,  g_{\mu \nu}~=~\partial_{a_j}  g_{\mu \nu} ~+~  Y_{(j)}^\rho \, \partial_\rho \, g_{\mu \nu}   ~+~ \partial_\mu \, Y_{(j)}^\rho \,  g_{\rho \nu}  ~+~ \partial_\nu \, Y_{(j)}^\rho \,  g_{ \mu \rho} \,, 
\label{CovVariations}
\end{equation}
where $\cL_{Y_{(j)}}$  denotes the Lie derivative.   The $Y_{(j)}^\rho$ are known as {\it compensating vector fields}, and are uniquely defined if one requires smoothness of $\delta_j g_{\mu \nu}$ and imposes the transverse, traceless gauge conditions:
\begin{equation}
	\nabla^\mu\, \delta_j   g_{\mu \nu}~=~ 0\,, \qquad  g^{\mu \nu} \, \delta_j   g_{\mu \nu}~=~ 0\,. 
	\label{TTmodes}
\end{equation}
Note, in particular, that the modes (\ref{LichModes}) are smooth and satisfy (\ref{TTmodes}).  This defines the exact relation between harmonic forms and variations of the metric with respect to moduli. 

It is convenient to define the covariant derivatives on the moduli space by setting: 
\begin{equation}
\cD_j ~\equiv~ \partial_{a_j} ~+~ \cL_{Y_{(j)}} \,. 
	\label{CovDer1}
\end{equation}
It is also important to note that the compensators transform as gauge fields under coordinate transformations that depend upon moduli.  That is, suppose that one makes a transformation from coordinates $x^\mu$ to coordinates $x^{\tilde \mu}  = x^{\tilde \mu} (x^\mu; a_j)$.  The covariant derivative in the new coordinates is given by
\begin{equation}
 \cD_j ~\equiv~ \partial_{a_j} ~+~ \cL_{\tilde Y_{(j)}} \,, 
	\label{CovDer2}
\end{equation}
where
\begin{equation}
\tilde Y_{(j)}^{\tilde \mu}  ~=~ Y_{(j)}^{ \mu}  \, \frac{\partial x^{\tilde \mu}}{\partial x^{ \mu}}  ~+~ \frac{\partial x^{\tilde \mu}}{\partial a_j}\bigg|_{x^\mu \ {\rm fixed} } \,.
	\label{gaugetrf}
\end{equation}
The second term on the right-hand side of (\ref{gaugetrf}) arises from the fact that $\partial_{a_j}$ in (\ref{CovDer1}) involves holding $x^\mu$ fixed while in (\ref{CovDer2}) it involves holding $x^{\tilde \mu}$ fixed.

The compensators for GH metrics in the coordinates $(\psi, \vec y)$ were computed in  \cite{Schulz:2012wu,Tyukov:2017jwn}.  When the GH points all lie along the $z$-axis, one has  
\begin{equation}
Y_{(j)}^{ \mu}  \,\frac{\partial}{\partial x^{ \mu} } ~=~ \frac{q_j}{r_j \, V} \, \partial_z \,.
	\label{comps1}
\end{equation}
The significant point is that $\frac{q_j}{r_j \, V}$ vanishes if $r_i =0$, for $i \ne j$, and limits to $1$ as $r_j \to 0$.  This means that  $\cD_j (r_k^{-n})$ is not more singular that the original function, $r_k^{-n}$, for all values of $k$.  One can also check  \cite{Tyukov:2017jwn} that 
\begin{equation}
\cD_j \, J   ~=~\Theta \,, \qquad  \cD_j \,g_{\mu \nu}  ~=~  \coeff{1}{2} \, \big( {J_\mu}^\rho \, \Theta_{\rho \nu} ~+~{J_\nu}^\rho  \, \Theta_{\rho \mu  } \big)   \,,
	\label{metvar1}
\end{equation}
where $\Theta$ is given by (\ref{harmtwoform}) and (\ref{Kharms}) with $k_i^{(I)} =0$ for $i \ne j$ and $k_j^{(I)} =1$

\subsection{Solving the second BPS equation via K\"ahler deformations }
\label{ss:NiceRes}

Consider the scalar Laplacian in a family of GH metrics, $g_{\mu\nu}(x^\rho; a)$, where the $a$ is a modulus, and suppose that one finds a family of  functions, $\widehat \Phi(x^\mu; a)$, that are harmonic in this metric.  Further suppose that we have chosen the compensators  so that 
\begin{equation}
\cD_a \, g_{\mu \nu}  ~=~  {J_\mu}^\rho \, \Theta_{\rho \nu}   ~=~    {J_\nu}^\rho \, \Theta_{\rho \mu}  \,.
	\label{metvar2}
\end{equation}
for some harmonic $2$-form, $\Theta$.  In particular, note that $\cD_a \, g_{\mu \nu}$ is traceless. This means that 
\begin{equation}
\cD_a \,g ~\equiv~ \cD_a \, \det(g_{\mu \nu})  ~=~  \det(g_{\mu \nu}) \, g^{\rho \sigma} \, \cD_a \, g_{\rho \sigma} ~=~ 0  \,.
	\label{covDrtg}
\end{equation}
If one writes the scalar Laplacian as
\begin{equation}
\nabla^2 \widehat \Phi ~=~  \frac{1}{\sqrt{g}} \, \partial_\mu \Big(\sqrt{g}\,  g^{\mu \nu}\,   \partial_\nu \widehat \Phi \Big) \,,
	\label{ScalLap}
\end{equation}
one can easily  see that 
\begin{equation}
\cD_a\, \nabla^2 \widehat \Phi ~=~  \frac{1}{\sqrt{g}} \, \partial_\mu \Big(\sqrt{g}\,  (\cD_a\,g^{\mu \nu})\,   \partial_\nu \widehat \Phi \Big) +  \frac{1}{\sqrt{g}} \, \partial_\mu \Big(\sqrt{g}\,  g^{\mu \nu}\,   \partial_\nu (\cD_a\,\widehat \Phi) \Big) \,,
	\label{varScalLap}
\end{equation}
Thus, if $\widehat\Phi$ is harmonic, one has
\begin{equation}
\nabla^2 \,(\cD_a\,\widehat \Phi)  ~=~  - \frac{1}{\sqrt{g}} \, \partial_\mu \Big(\sqrt{g}\,  (\cD_a\,g^{\mu \nu})\,   \partial_\nu \widehat\Phi \Big)~=~  - \nabla_\mu \big( (\cD_a\,g^{\mu \nu})\,   \partial_\nu \widehat\Phi \big)  \,.
	\label{Zstep1}
\end{equation}
However, the fact that $\cD_a\,(\delta_\mu^\nu) =0$ implies
\begin{equation}
\cD_a\,g^{\mu \nu} ~=~ - g^{\mu \rho} \, g^{\nu \sigma}\,(\cD_a\,g_{\rho \sigma} )~=~  -  \Theta^{\mu \rho} \, {J_{\rho}}^\nu    \,.
	\label{Dinvmet}
\end{equation}
Now assume that $\Theta$ has a pre-potential, $\Phi$, and use $\widehat \Phi$ to define a second, independent $2$-form  $\widehat \Theta$.  That is, define  $\Phi$ and $\widehat \Theta$ via
\begin{equation}
\Theta   ~=~    d\, \big( \big({J_\mu}^\rho \, \partial_ \rho \, \Phi \big) \,dx^\mu \big)\,, \qquad \widehat \Theta   ~=~    d\, \big( \big({J_\mu}^\rho \, \partial_ \rho \,\widehat \Phi \big) \,dx^\mu \big)  \,.
\label{Prepots2}
\end{equation}
One then has
\begin{equation}
\nabla^2 \,(\cD_a\,\widehat \Phi)  ~=~ \nabla_\mu \,\big({\Theta^{\mu \nu}}\, {J_{\nu }}^\rho\,   \partial_\rho \widehat\Phi \big) ~=~  {\Theta^{\mu \nu}}\,\nabla_\mu \,( {J_{\nu }}^\rho\,   \partial_\rho \widehat\Phi ) ~=~ \coeff{1}{2}\, \Theta^{\mu \nu}\, \widehat \Theta_{\mu \nu} ~=~ \star_4 \,\big( \Theta \wedge \widehat \Theta\big)\,,
	\label{Zstep2}
\end{equation}
where we have used the fact that the divergence of $\Theta$ vanishes because it is self-dual and closed.   We therefore see that $\cD_a\,\widehat \Phi$ provides a solution to the second layer of BPS equations,  (\ref{BPSeqn:2}).  

To summarize, suppose that a modulus, $a$, is associated with a harmonic $2$-form, $\Theta$, and that we find the pre-potential, $\widehat \Phi$, for another harmonic $2$-form, $\widehat \Theta$, at generic values of $a$.  Then the solution to the second BPS equation sourced by $\Theta \wedge \widehat \Theta$ is proportional to  $\cD_a\,\widehat \Phi$.

This has two important consequences.  First, if $a$ is one of the standard GH moduli, and  $\widehat \Phi$ is a pre-potential for a non-trivial fluctuating harmonic form of the class discovered in \cite{Bena:2017geu} then $\cD_a\,\widehat \Phi$ solves the BPS equation (\ref{BPSeqn:2}) for the interaction of the fluctuating $\widehat \Theta$ with the canonical GH fluxes of the form (\ref{harmtwoform}).  

Conversely, if $\widehat \Phi$ is a pre-potential for a non-trivial fluctuating harmonic form and $a$ is the putative modulus for a non-trivial fluctuating K\"ahler deformation, then solving the second BPS equation tells us how the more general moduli space develops at second order in such variations.  Interestingly, it is an empirical fact that {\it regularity} of the solutions to the BPS equations with sources that are second order in fluctuating modes requires the sources to satisfy local charge-density constraints  (see, for example,\cite{Bena:2010gg, Bena:2013ora,Bena:2014rea}).   One of the standard ways of addressing this constraint is through  ``coiffuring'' the modes   \cite{Bena:2013ora,Bena:2014rea,Bena:2015bea}.  In particular, the complex Fourier coefficients of fluctuating modes must  satisfy quadratic constraints if they are to lead to smooth solutions.  This suggests that generic deformations of the K\"ahler potential by fluctuating modes are obstructed at second order and that  regular deformations are restricted to sub-manifolds (defined via ``quadrics'') in the space of complex Fourier coefficients.

 
\section{Two-centered example}
\label{Sect:TwoCenter}

\subsection{The two-centered BPS solution}
\label{ss:2CenterSol}
To illustrate the general discussion of Section \ref{Sect:ComplexGH}, we apply the ideas developed there to the  five-dimensional supertrata solutions presented in \cite{Bena:2017geu}. The underlying supergravity theory is five-dimensional, $\Neql 2$ supergravity coupled to $n=3$ vector multiplets.  The indices, $I,J,K, \dots$ take values in $\{ 1,2,3,4\}$ and  the structure constants are given by:
\begin{align}
C_{123}=1\,, \qquad C_{344}=-2\,,
\end{align}
with all other independent components of $C_{IJK}$ vanishing.

The solutions in \cite{Bena:2017geu} are defined on an ambi-polar GH base with two centers:
\begin{align}
(q_{1},q_{2}) ~=~ (q_{-},q_{+})~=~(-1,+1) \,, \qquad  (a_{1},a_{2}) ~=~ (a_{-},a_{+}) ~=~ \left(-\frac{a^{2}}{8}, \frac{a^{2}}{8} \right) \,,
\end{align}
where $a$ provides a convenient parametrization of the single GH modulus of this geometry: the size of the $2$-cycle. We have also introduced subscript $\pm$ to label the distinct centers by their corresponding charges. Since the net charge is zero, the geometry of this solution will be asymptotically $AdS_{3}$.

The standard, non-fluctuating microstate geometry based on a single bubble in five-dimensions  is usually determined entirely in terms of the magnetic flux parameters.  The electric charges and angular momentum vector  are then fixed in terms of these fluxes.   The solutions of interest here have non-trivial, non-fluctuating parts of the $\Theta^{(I)}$, $I=1,2,3$, while $\Theta^{(4)}$ is purely fluctuating.  There are thus only three non-fluctuating flux parameters. Moreover, the original solution was obtained from spectral transformations and dimensional reduction of a D1-D5 system, which is most simply characterized in terms of the D-brane charges, $Q_{1},Q_{5}$, and the magnetic flux, $R\, \kappa$, of  $\Theta^{(3)}$.  Thus the fundamental, non-fluctuating parts of the solution will be characterized in terms of these three parameters\footnote{The factor of $R$ in the magnetic flux is related to the radius of the circle in the six-dimensional solution.}.

The fluctuating parts are governed by two Fourier coefficients: $b$ and $b_4$.  The former determines the fluctuations in the pair $(Z_1, \Theta^{(2)})$ while the latter determines the fluctuations of $(Z_4, \Theta^{(4)})$.  The pair, $(Z_2, \Theta^{(1)})$ has no fluctuating parts.  Finally, the function $Z_3$ and the angular momentum vector, $k$, are sourced quadratically in the  fluctuating modes.  Regularity then requires that $b$ and $b_4$ are proportional to one another and, ultimately,  only the RMS values of the fluctuations appear in the source terms for $Z_3$ and $k$.  Thus the third electric charge, $Q_3$, and the angular momentum components are determined in terms of RMS value of the fluctuations, $b^2$.

The solution is therefore parametrized by $(Q_{1},Q_{5},R\kappa,b,b_{4}, a)$.  However, it was shown in \cite{Bena:2016ypk, Bena:2017xbt} that regularity of these solutions requires a modified supertube relation and a relationship between the Fourier coefficients:  
\begin{equation}
Q_{1}Q_{5} = (R\kappa)^{2}\left( a^{2}+\frac{b^{2}}{2}\right)  \,, \qquad b_{4}^{2} = b^{2} \begin{pmatrix}
2m \\ m
\end{pmatrix} \begin{pmatrix}
2m+n-1 \\n
\end{pmatrix}\,.
\end{equation}
In the five-dimensional solutions this leads to regularity at the GH points. 

We now catalog the details of the complete five-dimensional solution given in \cite{Bena:2017geu}.

The two-centered solution is much easier to analyze because the wave equation is separable in  oblate spheroidal coordinates.\footnote{There are some orientation issues in translating between the standard GH coordinates and the spherical bipolar coordinates.  This is most conveniently resolved by sending $\phi \to -\phi$ in going between the two sets of coordinates. We present the solution in its spherical bipolar form.} So we begin by going over to these coordinates, $(r,\theta)$, defined via:
\begin{equation}
y_{1}+iy_{2}~=~  \frac{r}{4}\sqrt{r^{2}+a^{2}}\sin2\theta\,e^{i\phi} \,, \qquad  y_{3}~=~ z ~=~\frac{1}{8}(2\,r^{2}+a^{2})\cos2\theta\,.
\end{equation}
 where $\theta \in [0,\pi/2)$, $r\in[0,\infty)$ and $\phi\in[0,2\pi)$.
The canonical  complex coordinates of (\ref{z12defn}) are then:
\begin{equation}
 \zeta_{2} ~=~ \frac{r}{4}\sqrt{r^{2}+a^{2}} \, \sin 2\theta \,e^{-i\phi} \,,  \qquad \zeta_{1}~=~\frac{r^{2}}{r^{2}+a^{2}} \,e^{-i\psi} \,.
\end{equation}
Using these coordinates one can define
\begin{align}
\Sigma~=~4\, r_{-}~=~ r^{2}+a^{2}\cos^{2}\theta \,, \qquad \Lambda~=~4\, r_{+}~=~r^{2}+a^{2}\sin^{2}\theta \,,
\end{align} 
in terms of which one has
\begin{align}
V ~ =~  \frac{4(\Sigma -\Lambda)}{\Sigma \Lambda} \,, \qquad  A~=~-\frac{2a^{2}(\Sigma+\Lambda)}{\Lambda \Sigma} \sin^{2}\theta \cos^{2}\theta \, d\phi~.
\end{align}

The solutions to (\ref{BPSeqn:1})--(\ref{BPSeqn:3}) we wish to study have a clean decomposition into non-fluctuating and fluctuating pieces, corresponding to trivial and non-trivial functional dependence on the angular coordinates $(\psi,\phi)$ respectively. The fluctuating pieces will depend on two integers $(m,n)$ and generally involve the following function and moding
\begin{equation}
\Delta_{m,n}~=~ \frac{a^{2m}r^{n}}{2^{m}(r^{2}+a^{2})^{\frac{1}{2}(2m+n)}}\sin^{m}2\theta \,,\qquad  \chi_{m,n} = \frac{1}{2}(m+n)\psi-m\phi \,. \label{moding}
\end{equation}
When an object has both non-fluctuating and fluctuating components we will denote the former  with a bar and the latter  with a tilde, as in (\ref{oscbreak0}). 

The solution has a purely non-fluctuating $\Theta^{(3)}$:
\begin{equation}
 \Theta^{(3)}~=~d\beta \,, \qquad \beta ~\equiv~ \frac{R\,\kappa }{2\cos 2\theta} \left(\frac{2r^{2}+a^{2}}{2a^{2}}d\psi-d\phi \right) \,. 
\end{equation}
The first component equation in (\ref{BPSeqn:2}) can be broken up  into non-fluctuating and fluctuating pieces:
\begin{equation}
\nabla^{2} \overline{Z}_{1} ~=~ \star_{4}\left(\overline{\Theta}^{(2)}\wedge \Theta^{(3)} \right) \,,\qquad  \nabla^{2} \widetilde{Z}_{1} ~=~ \star_{4}\left(\widetilde{\Theta}^{(2)}\wedge \Theta^{(3)} \right)\,, \label{BPSex2}
\end{equation}
where
\begin{equation}
Z_{1}~=~ \overline{Z}_{1}+\widetilde{Z}_{1} \,, \qquad \Theta^{(2)} ~=~\overline{\Theta}^{(2)}+\widetilde{\Theta}^{(2)}\,. \label{oscbreak0}
\end{equation}
The second component equation in (\ref{BPSeqn:2}), involving $(Z_{2},\Theta^{(1)})$, has a similar decomposition, however, the solution in  \cite{Bena:2017geu}   has no fluctuating part in $(Z_{2},\Theta^{(1)})$.

The solutions of $\Theta$'s are given in \cite{Bena:2017geu} in terms of  auxiliary one forms $\lambda$.  In particular, one has 
\begin{equation}
\overline{\Theta}^{(I)} ~=~ (1+\star_{4})\left[(d\psi+A)\wedge \overline{\lambda}_{I} \right]\,, \qquad \widetilde{\Theta}^{(I)}  ~=~ (1+\star_{4})\left[(d\psi+A)\wedge \widetilde{\lambda}_{I} \right] \,, \quad I = 1,2, 4\,. \label{Thetadlam}
\end{equation}

One can check, by direct calculation, that the following is a solution to the first components of (\ref{BPSeqn:2}):
\begin{align}
\overline{Z}_{1}&~=~\frac{Q_{1}}{\Sigma-\Lambda}\,, \qquad   \widetilde{Z}_{1}~=~\frac{Q_{1}}{\Sigma - \Lambda}\left(\frac{b_{4}^{2}}{2a^{2}+b^{2}} \right) \Delta_{2m,2n}\cos \chi_{2m,2n} \,, \label{Z1sol} \\
\overline{\lambda}_{2}&~=~ - \frac{1}{4  R \kappa}d_{3}\left[ (\Sigma+\Lambda) \overline{Z}_{1}\right] \,, \\
 \widetilde{\lambda}_{2}&~=~- \frac{1}{4 R\kappa}d_{3}\left[(\Sigma+\Lambda)\widetilde{Z}_{1} \right] + \frac{a^{2}(m+n)}{4R\kappa}\left[4 \widetilde{Z}_{1}\cot 2\theta \, d\theta +\frac{1}{m}\partial_{\phi}\widetilde{Z}_{1}\, d\phi\right]\,,
\end{align}
where $d_{3}$ is the exterior derivative on the $\IR^3$ base of GH metric, described by the coordinates $\vec{y}$.

The pair $(Z_{2},\Theta^{(1)})$ is non-fluctuating and is given by  (\ref{Thetadlam}) and
\begin{equation}
Z_{2} ~=~ \frac{Q_{5}}{\Sigma-\Lambda} \,, \qquad  \lambda_{1} ~=~ -\frac{1}{4R\kappa}  \, d_{3} \left[(\Sigma+\Lambda)Z_{2} \right]   \label{Z2Th1sol}  \,.
\end{equation}
The pair $(Z_{4},\Theta^{(4)})$ is purely fluctuating and is given by:
\begin{equation}
\begin{aligned}
Z_{4}&= -\frac{2b_{4}R\kappa}{\Sigma-\Lambda}\Delta_{m,n}\cos\chi_{m,n}   \,, \\
\lambda_{4} &= \frac{1}{8R\kappa} \left[ d_{3} \left[(\Sigma+\Lambda)Z_{4} \right] -a^{2}(m+n)\left(2Z_{4}\cot 2\theta \, d\theta +\frac{1}{m}\partial_{\phi}Z_{4}\, d\phi \right) \right]  \,.
\end{aligned}
\label{Z4Th4sol} 
\end{equation}

The remaining component of (\ref{BPSeqn:2}) leads to the following equation for  $Z_3$:
\begin{equation}
\nabla^{2} Z_{3} ~=~ \star_{4}\left(\Theta^{(1)}\wedge \Theta^{(2)}-\Theta^{(4)}\wedge \Theta^{(4)} \right) \,.
\end{equation}
and the last equation, (\ref{BPSeqn:3}), is solved by decomposing $k$ into a component, $\mu$, along the GH fiber  and 1-form, $\varpi$, on the $\IR^3$ base of the GH metric.   The analytic solutions for arbitrary $(m,n)$ are not known in closed form, but we can summarize the infinite family of solutions for $m=1$ and arbitrary $n$:
\begin{equation}
Z_{3}~=~ \frac{1}{2(\Sigma-\Lambda)}\left(2a^{2}+b^{2}-\frac{b^{2}_{4}}{\sin^{2}2\theta}\Delta_{2,2n} \right) \,, \qquad k=\mu(d\psi +A)+\varpi \,,
\end{equation}
with 
\begin{align}
\mu&~=~ \frac{R\kappa}{8a^{4}} \left[ \frac{2b^{2}r^{2}\Delta_{1,2n}}{\sin 2\theta} -a^{2}(2a^{2}+b^{2})+\frac{a^{2}+2r^{2}}{\cos^{2}2\theta}\left( (2a^{2}+b^{2})-b_{4}^{2}\Delta_{2,2n}\right) \right]\,,\\
\varpi &~=~ \frac{R\kappa}{16 \Sigma \Lambda} \left[ \frac{2r^{2}\left( a^{2}b_{4}^{2}+b^{2}(a^{2}+2r^{2})\right)}{a^{2}\sin 2\theta}\Delta_{1,2n} -2r^{2}(2a^{2}+b^{2}) \right] \sin^{2}2\theta \, d\phi \,.
\end{align}

This summarizes the complete five-dimensional superstratum solution given in \cite{Bena:2017geu}.

\subsection{Pre-potentials, compensator and deformations}
\label{ss:PrepotentialsOscSol}

As a non-trivial application of the technology developed in Section \ref{Sect:ComplexGH}, one can directly verify that there are indeed pre-potentials from which $\overline{\Theta}_{2} $ and $\widetilde{\Theta}_{2} $ are constructed:
\begin{align}
\overline{\Phi}_{2} &= -\frac{Q_{1}}{2R\kappa} \ln ( \tan\theta)+C_{1}\ln \left|\zeta_{1} \right|+C_{2}\ln \left|\zeta_{2} \right|+C_{3}\,, \label{prepot1}\\
\widetilde{\Phi}_{2} &= \frac{1}{2 R\kappa} \Re\left\lbrace \frac{\zeta^{m+n}_{1}}{\zeta^{2m}_{2}} \right\rbrace \left(\frac{a^{4m}b_{4}^{2}Q_{1}}{2^{6m}(2a^{2}+b^{2})} \right)\left[C_{4}+\,    _2F_1 \left(\frac{1}{2},1-2m ,\frac{3}{2},\cos^{2}2\theta\right) \cos 2\theta \right] \,, \label{prepot2}
\end{align}
where $_{2}F_{1}$ is the ordinary hypergeometric function, and 
\begin{align}
\left|\zeta_{1} \right| ~&=~ \frac{r^{2}}{r^{2}+a^{2}} \,, \qquad   \left|\zeta_{2} \right|~=~ \frac{r}{4}\sqrt{r^{2}+a^{2}}\,\sin 2\theta \,, \\
  \Re\left\lbrace \frac{\zeta^{m+n}_{1}}{\zeta^{2m}_{2}} \right\rbrace~&=~ \frac{2^{4m}r^{2n}}{(r^{2}+a^{2})^{2m+n}\sin^{2m}2\theta} \cos \chi_{2m,2n} \,.
\end{align}
We have included the homogeneous pieces involving the constants $(C_{1},C_{2},C_{3},C_{4})$.  These terms give a vanishing contribution when substituted in (\ref{Prepot}) because the derivatives in (\ref{Prepot}) are equivalent to taking $\partial_{\zeta_i}  \partial_{\bar{\zeta}_j}$.  While these homogeneous terms are ``pure gauge,'' they are important in the singularity analysis of Section \ref{ss:SingularStructure}.

The compensator can be directly calculated from equation (\ref{metvar1}) or equation (\ref{comps1}). Since we are  using $a$ as a modulus, it is more convenient to replace $\partial_{a_{j}}$ by $a\partial_{a}$ in the definition of $\mathcal{D}_{a}$.  The extra factor of $a$ has been included to make $\mathcal{D}_{a}$ dimensionless. A simple calculation shows that the corresponding dimensionless compensator in $\mathcal{D}_{a}$ is given by the vector field
\begin{equation}
Y = r\,\partial_{r} - \tan 2\theta \,\partial_{\theta} \,. 
\end{equation}
With the compensator and pre-potentials in hand, we can check that (\ref{Zstep2}) does indeed give (\ref{Z1sol}). Indeed, we find:
\begin{equation}
\overline{Z}_{1}~=~  \frac{R\kappa}{a^{2}}\,\mathcal{D}_{a}\overline{\Phi}_{2}    \,, \qquad  \widetilde{Z}_{1}   ~=~ \frac{R\kappa}{a^{2}} \frac{a^{4m}}{2a^{2}+b^{2}} \, \mathcal{D}_{a} \left(\frac{2a^{2}+b^{2}}{a^{4m}}\widetilde{\Phi}_{2} \right) \,. \label{PhiToZ}
\end{equation}

In general $\mathcal{D}_{a} \Phi$, will only provide a particular solution for  $Z$ and one may need to add homogeneous solutions to arrive at the suitably regular potential.  Such homogeneous solutions are harmonic and are thus completely determined by their boundary conditions and singular structure.  Given that $\Phi$ has a prescribed singularity structure, the necessary homogeneous solutions are generically going to be proportional to $\Phi$ itself.  Indeed, in (\ref{PhiToZ}), we have inserted a judiciously chosen factor of $a^{-4m} (2a^{2}+b^{2})$ inside the derivative so as to generate precisely $\widetilde{Z}_{1}$ without the need to add some multiple of $\Phi$.

\subsection{Singular structure}
\label{ss:SingularStructure}

We now return to the issue of the singular structure of the pre-potentials mentioned at the end of Section \ref{ss:harmonic}, and expand upon those comments based on our example. Given that the harmonic 2-forms, $\Theta^{(I)}$, are permitted to be singular on the $V=0$ ``evanescent ergosurface"
\begin{equation}
V=0 \quad   \Leftrightarrow\quad \theta = \frac{\pi}{4} \quad\Leftrightarrow \quad z=0   \,,
\end{equation}
one might expect the pre-potentials of (\ref{prepot1})--(\ref{prepot2}) also to share this property. However, we find something very different: the harmonic functions are singular at $\theta=0$ or $\theta=\pi/2$, or both. 

In $\widetilde{\Phi}_{2}$ the singularities arise due to the $\sin^{-2m}2\theta$ factor since the hypergeometric term is regular at these points. In fact one finds
\begin{equation}
\left. \cos (2\theta) ~ _2 F_1 \left(\frac{1}{2},1-2m,\frac{3}{2},\cos^{2}2\theta \right) \right|_{\theta=0} =  -\left. \cos (2\theta) ~ _2 F_1 \left(\frac{1}{2},1-2m,\frac{3}{2},\cos^{2}2\theta \right) \right|_{\theta=\pi/2} ~=~  K_{m} \,, \label{GammaCon}
\end{equation}
where
\begin{equation*}
K_{m}~\equiv~ \frac{\sqrt{\pi}\,\Gamma (2m)}{2\Gamma\left( \frac{1}{2}+2m\right)}\,.
\end{equation*}
Hence, when one expands about $\theta=0$ or $\theta=\pi/2$, up to overall constants the singular terms take the form 
\begin{align}
\widetilde{\Phi}_{2} & ~\sim~ \frac{1}{\theta^{k}} \left( \frac{a^{2}}{2^{2}(a^{2}+r^{2})}\right)^{2m} \left( \frac{r^{2}}{a^{2}+r^{2}}\right)^{n} \left(C_{4} +K_{m}\right)\,,\\
\widetilde{\Phi}_{2} & ~\sim~ \frac{1}{\left(\theta- \frac{\pi}{2}\right)^{k}} \left( \frac{a^{2}}{2^{2}(a^{2}+r^{2})}\right)^{2m} \left( \frac{r^{2}}{a^{2}+r^{2}}\right)^{n} \left(C_{4} -K_{m} \right)\,,
\end{align} 
where $k\leq 2m$ is an even integer. 

On the face of it, choosing  $C_{4} =\mp K_{m}$ will remove the  leading singularity in  $\widetilde{\Phi}_{2}$.  However, the remarkable thing is that these choices remove the singularities {\it to all orders} for $\theta=0$ and $\theta=\pi/2$ respectively.   

This seemingly remarkable property has a very simple explanation in the form of the identity:
\begin{equation}
x \, \left[  _2 F_1 \left(\frac{1}{2},-p,\frac{3}{2}, x^2 \right) \right]  ~= ~  \int \, (1-x^2)^{p} \, dx \,.
\label{HypGeoId}
\end{equation}
Thus, if one chooses the constant of integration correctly, this expression has a zero of order $(p+1)$ at either $x=+1$ or $x=-1$.

Now recall that  $C_4$ multiplies the real part of a holomorphic function and so it is pure gauge in that it disappears when one constructs the flux, $\Theta$.  Therefore the location of the singularity is simply a gauge choice and can be flipped between the ambi-polar patches where either $V>0$ or $V<0$.   Moreover, the fact that these singularities must be pure gauge determines the singular structure of $\Phi$: the only singularities appearing in $\widetilde{\Phi}_{2}$ have the same form as purely holomorphic gauge terms.  
 
Looking back at (\ref{Prepots2}) ,we see that the singularities in the fluxes turn up on the $V=0$ surface implicitly through the complex structure, $J$. Using this insight one might hope to generalize this procedure: For a GH base manifold with $n$ co-linear centers one might  obtain new and interesting fluxes by first constructing harmonic functions that are everywhere regular except on the line through the GH points, where they have singularities that are precisely defined by the vanishing of some holomorphic function.  In this way, the singularities of the pre-potential will disappear because they  gauge artifacts when used in  (\ref{Prepot}) to obtain the harmonic $2$-forms. In the next section we will comment further on the interpretation of the singularities of the pre-potentials.

\section{Discussion and speculation}
\label{Sect:Conc}

We have used the K\"ahler structure, and K\"ahler deformation theory, to simplify and illuminate  the structure of the first two BPS equations that govern microstate geometries in five dimensions.   

There are several motivations for this work.  In physics, we wish to push beyond the well-explored ambi-polar, two-centered solutions and find fluctuating microstate geometries for multi-centered metrics.  For two centers, the analysis is greatly simplified because of the separability of the wave-equation on the GH base, $\cB$.  Moreover, the two-centered geometries can be related to a single supertube in flat space and from this one can construct many of the scalar and tensor harmonics using rigid $\cR$-symmetry rotations. (See, for example, \cite{Mathur:2003hj,Giusto:2013bda,Shigemori:2013lta,Bena:2015bea}.)

For multi-centered geometries, the BPS equations are extremely challenging, but far from hopeless.  The Green function for the scalar Laplacian is known \cite{Page:1979ga} and its structure can be extended to ambi-polar spaces \cite{Bena:2010gg}.  This, at least, affords the possibility of generating families of scalar harmonics by expanding the Green function in terms of the source location.  Thus scalar harmonic analysis for general ambi-polar, multi-centered metrics is, in principle, within reach.  The fact that we have largely reduced the first two layers of the BPS equations to {\it harmonic, scalar} pre-potentials is therefore a very significant step towards the construction of multi-centered, fluctuating BPS microstate geometries.

More broadly, understanding fluctuating, multi-centered  microstate geometries is particularly appealing from the perspective of holographic field theory.   If one focusses on a single bubble of a microstate geometry then that bubble may be viewed as the dual of a  smaller CFT whose central charge is determined by the fluxes through the bubble.  Fluctuations of an individual bubble then represent excitations of the smaller CFT.  Thus a multi-bubble solution should represent a flow from a larger CFT in the UV to a product of smaller CFT's in the IR (as represented by the bubbles).  The holography of individual fluctuating bubbles would thus inform one about how states in the product theory are related to states in the large UV fixed-point theory.  One would then have  a raft of examples of holographic flows from one ``master CFT'' to a rich range of product CFT's in the IR.  This would also provide much-needed insight into the role of multi-centered geometries in describing the microstate structure of black holes.

There are several important ways in which one might think about extending this work.  First, and most obvious, is whether the third, and last, BPS equation for the angular momentum vector, $k$, also has some natural interpretation and solution in terms of the structure of K\"ahler deformations. It is well-known  \cite{Bena:2007kg,Bena:2010gg} that the component, $\mu$, of  $k$ along the fiber is determined by a scalar wave equation with a non-trivial source.  This, once again, makes it the easiest object to determine and characterize, but, thus far, we have not found a simple characterization  in terms of the structures discussed in this paper.  The source terms of the third BPS equations strongly suggest that  $k$ should play a role in the higher-order perturbation theory of the K\"ahler structure.  Moreover, the   vector, $k$, plays an absolutely essential role in the regularity of both the metric  (\ref{reg4met})  and the electromagnetic fields (\ref{regMax}), and so it should have  a precise and natural interpretation in the geometry of ambi-polar metrics.

Since ambi-polar metrics also play a major role as base metrics of six-dimensional BPS geometries, it would be very interesting to see if the six-dimensional BPS equations  also have pre-potentials and are related to deformations of the K\"ahler structure.  The most general four-dimensional base of a six-dimensional BPS geometry is ``almost hyper-K\"ahler" \cite{Gutowski:2003rg} and since these geometries do not appear to have a well-developed mathematical structure, it seems premature to work at such a level of generality.  On the other hand, there are large families of interesting, intrinsically six-dimensional BPS geometries that are based on GH base manifolds but with electromagnetic fields and warp factors that depend on all the spatial coordinates, and particularly upon the compactification circles that would otherwise enable one to compactify the solution down to five dimensions.  For this class of BPS geometries, the BPS equations may well admit a generalization of the results presented in this paper and we are currently investigating this. 

There are  further  questions and conjectures that arise from our results. First, we motivated the idea of pre-potentials for the fluctuating harmonic forms of ambi-polar geometries by drawing parallels with Riemannian geometries.  The fact that we found such pre-potentials means that, at least to first order, they also represent  hyper-K\"ahler deformations of the ambi-polar metrics\footnote{Such pre-potentials yield, to first order, Ricci-flat,  K\"ahler metrics, which, in four dimensions, means that they are actually hyper-K\"ahler.}.  {\it A priori}, it is not clear whether these perturbations are integrable to macroscopic moduli, or whether they are obstructed at some higher order.   The work on microstate geometries over the last few years gives some very interesting insights into how this story might evolve, and this might lead to a mathematical interpretation of the local supertube charge-density constraints, and the technique known as ``coiffuring.'' 

To understand this comment in more detail, first note that we have shown how the pre-potentials can be used to solve the second layer of the BPS equations when a fluctuating mode interacts with a ``canonical cohomological flux'' associated with a GH modulus.  While this is technically second-order in K\"ahler deformations, it is actually first order in the canonical GH modulus and first order in the Fourier coefficients of the fluctuating modes.  To solve the complete second layer of BPS equations one needs to solve (\ref{BPSeqn:2}) when both fluxes are fluctuating, which means one is working at second order in the Fourier coefficients of fluctuating modes.  Thus completely solving the second layer, (\ref{BPSeqn:2}),  of the BPS system is pushing these putative new K\"ahler moduli to second order in perturbation theory.  Similarly, solving the last layer of the BPS system, (\ref{BPSeqn:3}), involves deformations to third order.  However, because of the form of the solutions that have been constructed so far, the source terms in (\ref{BPSeqn:3}) are limited to quadratic order in the  Fourier coefficients of fluctuating modes.    

Extensive computations of examples of fluctuating microstate geometries suggest that generic second-order perturbations lead to singularities in the fields and in the geometries  \cite{Bena:2010gg, Bena:2015bea,Bena:2016ypk,Bena:2017xbt}.  However, such singularities can be removed by solving a constraint on fluctuating charge densities \cite{Bena:2010gg}.    The technique of ``coiffuring'' was evolved \cite{ Bena:2013ora,Bena:2014rea,Bena:2015bea,Bena:2016ypk,Bena:2017xbt}  to address this issue when working with Fourier series with a finite number of modes.  In practice, what this means is that, when generic fluctuations are present, one must apply quadratic constraints to the Fourier coefficients in order to cancel singularities.  As a result of this, we now know that there are still  infinite families of smooth solutions with fluctuating Maxwell fields and fluctuating geometries \cite{Bena:2010gg, Bena:2015bea,Bena:2016ypk,Bena:2017xbt}, however the Fourier modes are constrained. It was also argued in \cite{Bena:2010gg, Bena:2015bea,Bena:2017xbt} that because the BPS equations are sourced by quadratic interactions of fluctuating modes, then, once one solves the quadratic constraints on the Fourier coefficients,  there are no further obstructions to solving the BPS system so as to obtain a smooth microstate geometry.   This suggests that generic K\"ahler deformations involving the  Fourier modes can be obstructed at second order but that there are unobstructed families when one combines some of the deformations in an appropriate manner.  

If the Fourier coefficients of the fluctuating modes can be integrated up to produce continuous families of hyper-K\"ahler deformations of ambi-polar geometries, then one would very much like to know what they are and how they can be characterized.  First, we note that because  the pre-potentials, and thus the families of metrics, depend non-trivially on the fiber, $\psi$,  there is no longer a tri-holomorphic $U(1)$ action.  If the solution still has some other $U(1)$ symmetry, then, in Riemannian geometry,  this would lead us to metrics that are characterized by solutions to $SU(\infty)$ Toda rather than the harmonic function, $V$, of GH metrics.  On the other hand, if one is optimistic, one might hope that, for ambi-polar families that connect smoothly to GH metrics, some of the simplicity of a metric based on harmonic functions would persist. 

One important clue as to the geometry underlying new families of ambi-polar metrics is the singular structure of the pre-potential.  At the outset of this work, we expected that the pre-potentials would have their singularities on the evanescent ergospheres simply because this is the natural singular locus.  Instead, we found that the pre-potentials have poles on divisors and the singularities emerge from real parts of holomorphic functions.  (The singular loci found in Section \ref{ss:PrepotentialsOscSol}  are defined by $\zeta_2 \zeta_1^{-1/2} =0$.)   The singular behavior is unphysical, or ``pure gauge,'' because the fluxes  are obtained by taking mixed derivatives, $\partial_{\zeta_i} \partial_{\bar \zeta_j}$, of the pre-potential.  Moreover, we saw in Section \ref{ss:PrepotentialsOscSol} that we could move the singularity of the pre-potential from one side of the evanescent ergosurface to the other simply by choosing different constants of integration.   This is very reminiscent of transition functions in non-trivial holomorphic vector bundles, in  that such transition functions are well behaved on the overlaps of patches by cannot be extended smoothly to the entire manifold\footnote{We are grateful to Emil Martinec for making this observation.}.  The data from these transition functions also precisely characterizes the topology underlying the bundles and thus one might expect something similar in ambi-polar topology.  

Finally, we cannot resist advancing a speculative idea about the mathematical structure of  harmonic  2-forms on ambi-polar spaces. For GH spaces with $N$ source points, the homology is $(N-1)$-dimensional and, in a canonical basis, the intersection matrix is the Cartan matrix of the Lie algebra $A_{N-1}$.  Indeed, if one puts all the GH points in general position, the cycles obtained by joining up the points in all possible ways, and with each possible orientation, can be associated with the roots of $A_{N-1}$.  The monodromies of fluxes and cycles as the GH points are moved can then be described through the  action of the Weyl group of $A_{N-1}$. The question is, how this might generalize to ambi-polar spaces. 

The intersection numbers of ambi-polar metrics were discussed in \cite{Gibbons:2013tqa} and the form very much depended on the ordering of the positively and negatively charged GH points.  Suppose there are $n_\pm$  GH points of charges $\pm 1$ and that we put them all along the $z$-axis with all the positive charges with $z<0$ and all the negative charges with $z>0$.  In the region $z<0$, the space is precisely that of a Riemannian GH metric with $n_+$ points and so the intersection matrix in this region is the Cartan matrix of $A_{n_+ -1}$.  Similarly, in the region $z>0$, the metric is the negative of  a Riemannian GH metric with $n_-$ points and the intersections of cycles are the same as a Riemannian GH space but with one exception.  Since the volume form on an ambi-polar GH space has a factor of $V$, it changes sign as one crosses an evanescent ergosurface.  This means that the intersection matrix for $z>0$ receives an overall change of sign relative to  $z<0$.  So the intersection matrix is the negative of the Cartan matrix of $A_{n_- -1}$.  As observed in  \cite{Gibbons:2013tqa}, the cycle crossing the evanescent ergosurface has a self-intersection of zero but  has intersection number $\mp 1$ with its neighbors, where the sign depends  on whether the signature of the metric is $\pm 4$ in the region around the intersection.  This describes the symmetric (as opposed to the ``distinguished'') Cartan matrix of the superalgebra $A(n_+,n_-)$ \cite{Frappat:1996pb}.  It is also interesting to note that the sensitivity of the intersection matrix to the ordering of the positive and negative GH points is reminiscent of the lack of a canonical choice of simple roots for a superalgebra. 

This raises the further question of whether the monodromies and fluxes on ambi-polar spaces are somehow related to the action of the Weyl group of $A(n_+,n_-)$, or some other  exotic discrete group.  The fact that we now have an infinite family of harmonic forms is certainly suggestive of a much larger monodromy group with elements of infinite order.  Indeed, the results in Section \ref{ss:2CenterSol} suggest that $\psi$-quantum number is something like an energy in that fixing this quantum number leads to finite-dimensional spaces of modes.  It would be extremely interesting to study this in more detail and, in particular, study  new examples in multi-centered ambi-polar geometries. 

This brings us full circle to the physical motivations for this work.  Our goal is ultimately to enable the explicit construction of fluctuating, multi-centered microstate geometries and classify the dual holographic states.  The results presented here make a  significant step in advancing this agenda by reducing some of the BPS equations to solving the scalar wave equation. We are currently adapting the results of \cite{Page:1979ga} to provide generating functions for scalar modes, with appropriate singular structure, on  multi-centered ambi-polar geometries and we hope to complete this work in the near future.  We also believe that we have uncovered what appears to be a very rich mathematical in the cohomology of ambi-polar geometries and that obtaining a deeper understanding of this will not only greatly streamline the analysis of the BPS equations but may well lead to whole new classes of fluctuating, ambi-polar, hyper-K\"ahler metrics.

\section*{Acknowledgments}

\vspace{-2mm}
We would like to thank Iosif Bena and Emil Martinec for valuable discussions.
RW and NPW would like to thank  the IPhT, CEA-Saclay for
hospitality while a some of this work was done.
This work  was supported in part by the DOE grant DE-SC0011687.





\end{document}